\documentclass[12pt,twoside]{article}
\usepackage{correl}
\usepackage{grffile}
\def\beq{\begin{equation}}
\def\eeq{\end{equation}}
\def\be{\begin{equation}}
\def\ee{\end{equation}}

\def\cG0{{\cal G}_0}

\def\spinup{\uparrow}
\def\spindown{\downarrow}

\def\D{\Delta}

\def\eps{\epsilon}

\def\s{\sigma}
\def\uc2{$U_{c2}$}
\def\uc1{$U_{c1}$}

\def\bea{\begin{eqnarray}}
\def\eea{\end{eqnarray}}
\def \bal{\begin{align}}
\def \eal{\end{align}} %%aliases don't work in these environments, see ams FAQ
\def\#{\!\!}
\def\@{\!\!\!\!}

\def\+{\dagger}

\def\up{\spinup}
\def\down{\spindown}

\def\TheTitle{Hund's metals, explained}
\def\TheHeading{Hund's metals}
\def\TheAuthor{Luca de' Medici}
\def\TheInstitute{Ecole Sup\'erieure de Physique et Chimie Industrielles de la Ville de Paris}
\def\TheAddress{10 rue Vauquelin, 75005 Paris, France}
\def\TheChapter{5}                       % Chapter (will be assigned by us)
\begin{document}
\MakeTitle           % displays title and table of contents
%%%%%%%%%%%%% put your text here:
\section{Introduction}
The expression "Hund's metal"\index{Hund's metal, definition} - to the best of the author's knowledge - was introduced in Ref. \cite{Yin_kinetic_frustration_allFeSC}, in the context of an ab-initio study of Fe-based superconductors (FeSC)\index{Fe-based superconductors}. It has met, since then, quite some success\cite{Hirjibehedin_NatNano_News,Kondo_OSM_HundsMetal_SrRu2O4_surface} and today is commonly used, without a complete consensus on its precise meaning. A phase in which Hund's coupling\index{Hund's coupling} - the intra-atomic exchange energy - influences crucially the metallic properties is probably a definition that encompasses the different current uses of this buzzword. 

In this chapter we will use a more specific - and hopefully precise - definition. 

Indeed we will show that, due to Hund's coupling, a metallic phase with specific features arises consistently in realistic simulations of paramagnetic FeSC, in accord with experiments where this has been tested (see Section \ref{sec:abinitio-exp}). This phase is found beyond a crossover line in the doping/interaction strength plane in the space of parameters, and is characterized by three main features, compared to the "normal" metal realized before the same frontier: 
\begin{enumerate}
\item enhanced electron correlations and masses,\\ 
\item high local spin configurations dominating the paramagnetic fluctuations,\\
\item selectivity of the electron correlation strength depending on the orbital character.\\
\end{enumerate}
These features grow both with increasing further the interaction strength and with proximity to the half-filling of the conduction bands (thus with hole-doping in FeSC), where a Hund's coupling favored Mott insulator is realized.

We will call this phase a Hund's metal.

In Section \ref{sec:models} we will show that these features are not specificities of the band structure of FeSC, because they arise consistently in models with featureless simplified densities of states. They are, in fact, outcomes of the many-body physics dictated by the local electronic configurations of the atomic orbitals that give rise to the conduction bands (the five orbitals of main Fe 3d character, in the case of FeSC). These configurations are shaped by the local coulomb interaction, and in particular by the Hund's coupling\cite{georges_annrev}, as it is known from atomic physics.

The role in this physics of the fermiology, and more in general of the specific features of the bare band structure in each particular case is comparatively lesser, and typically lumped in a few local parameters: the orbital energies, the total and orbitally-resolved kinetic energy, etc.
This reduced influence implies the generality of this physics and its relevance for many materials besides FeSC, which can be called Hund's metals. A recently very investigated example is that of Ruthenates\cite{Mravlje_SrRuO4_coherence,Kondo_OSM_HundsMetal_SrRu2O4_surface}\index{Ruthenates}.

In Section \ref{sec:analytics} we will then provide through analytical arguments some insight in the basic mechanisms by which Hund's coupling induces the above mentioned features.
 
Finally in Section \ref{sec:compress} we will outline a recent additional feature of Hund's metals, stemming plausibly from the same mechanisms outlined in the previous section: the enhancement (culminating in a divergence) of the electronic compressibility\index{electronic compressibility} in proximity of the crossover between the Hund's metal and the normal metallic phase. The divergence of the compressibility signals an instability towards phase separation/charge-density waves. Its enhancement signals enhanced quasiparticle interactions that can also favor instabilities. Both possibly link the Hund's metal crossover to high-Tc superconductivity.

\section{Hund's metals in Fe-based superconductors: experimental evidences and \emph{ab-initio} studies}\label{sec:abinitio-exp}

Many different families of Iron-based superconductors have been synthesized\cite{PaglioneGreene_NatRev}, all bearing as a central unit a buckled plane of Fe atoms disposed in a square array with ligands (As, P, Se or Te) in the middle of each square, alternatively slightly above or below the plane.
The so-called "122" family of FeSC (where FeAs layers are interleaved with buffer layers of Ba or other alkaline earth or alkali elements) is particularly well suited to highlight the Hund's metal phenomenology, as defined above. Indeed these compounds can be synthesized in high-quality single-crystals and cleaved easily to yield clean surfaces, all of which facilitates several experimental techniques, such as angular-resolved photoemission spectroscopy (ARPES) for example. 
As importantly, the mother compound BaFe$_2$As$_2$ (the chemical formula giving the name to the family) can be engineered through many chemical substitutions, that allow to tune continuously both the structural and the electronic properties so to map out finely a complex phase diagram. There, the gross features are: a high-temperature metallic paramagnetic phase that has tetragonal symmetry, becoming below some temperature either a tetragonal superconducting phase or an orthorhombically distorted magnetic phase, depending on the exact composition\footnote{with some exceptions, in some very small areas of the phase diagram: e.g. the two low-temperature phases can actually coexist, or a non-magnetic distorted (nematic) phase can be realized}.

\begin{figure}[t!]
 \centering
 \includegraphics[width=0.6\textwidth]{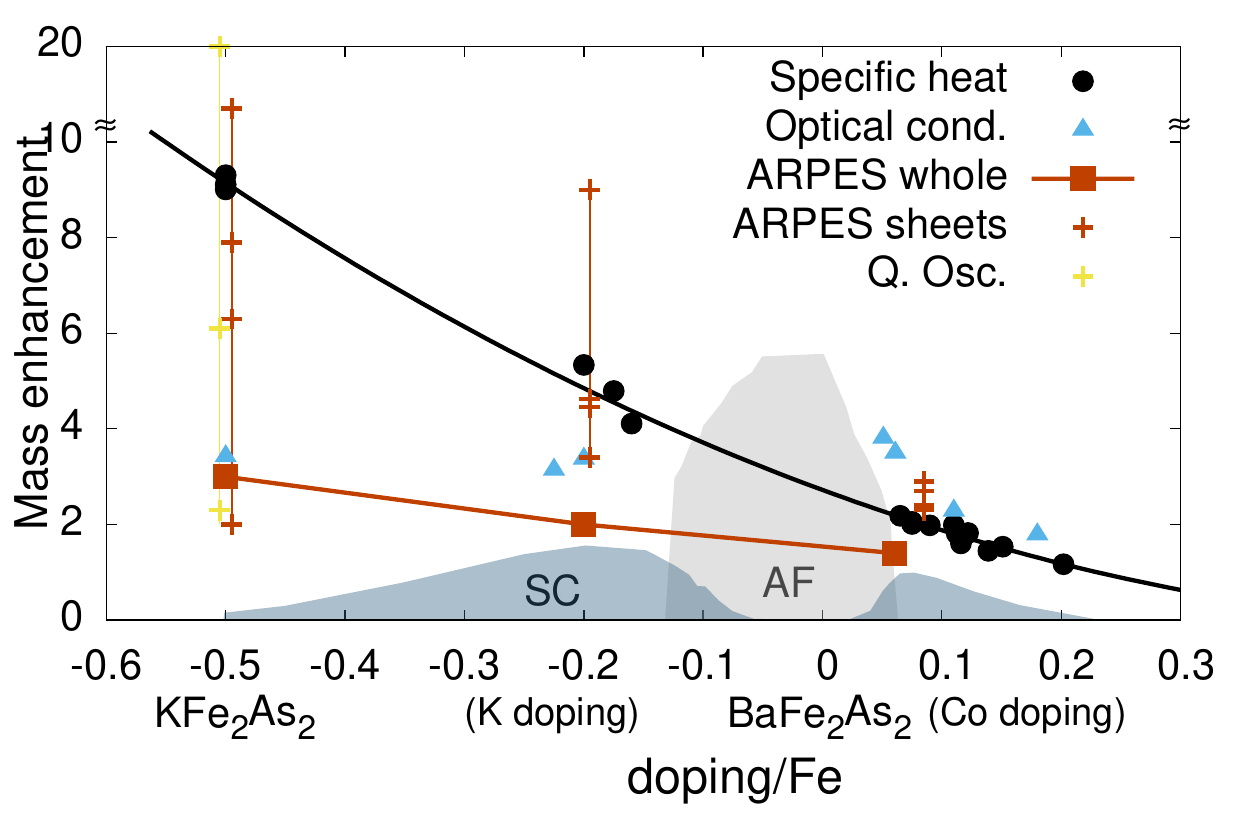}
 \includegraphics[width=0.6\textwidth]{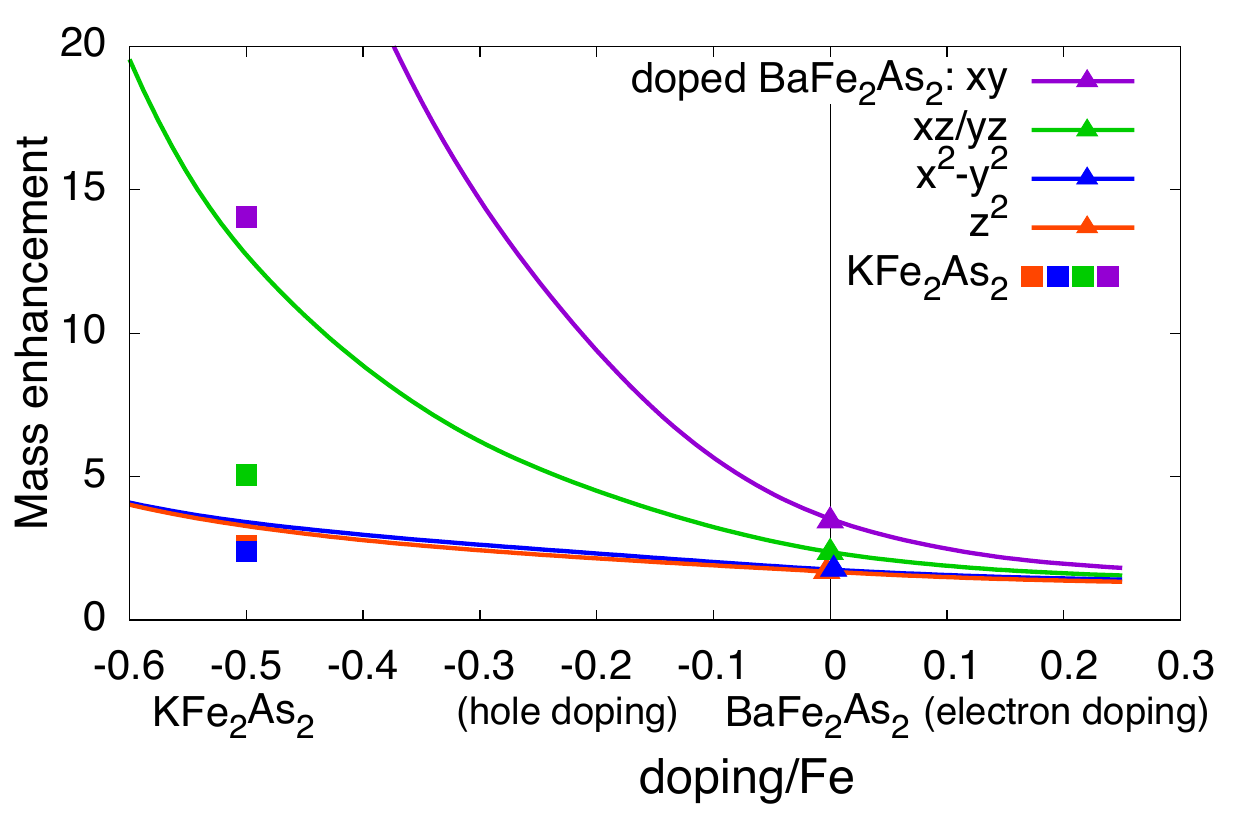}
 \caption{Upper panel: experimental estimates of the mass enhancement in the 122 family from the various probes indicated in the legend. As detailed in the main text the divergence of the estimates between the various band-integrated techniques can be interpreted in terms of coexisting weakly and strongly correlated electrons, and this is confirmed by the spread of the band-resolved values of ARPES and quantum oscillations. Lower panel: orbitally-resolved mass enhancement calculated within DFT+SSMF capturing the above mentioned behavior (adapted from Ref. \cite{deMedici_OSM_FeSC}).}
 \label{fig:FeSC_OSM}
\end{figure}

Pure BaFe$_2$As$_2$ becomes distorted and magnetic below $\sim$140K, whereas both electron doping (most commonly by partially substituting Fe with Co) and hole doping (substituting Ba with K) lead to the suppression of this phase and the rise of superconductivity (reaching a maximum T$_c\sim$23K for Ba(Fe$_{0.93}$Co$_{0.07}$)$_2$As$_2$ and T$_c\sim$38K for  Ba$_{0.6}$K$_{0.4}$Fe$_2$As$_2$, respectively). The hole doping can be continued until reaching another stoichiometric compound KFe$_2$As$_2$ (where T$_c\sim$3K).
This is the most extended doping range that can be continuously obtained in a single family of FeSC to date (the phase diagram on this range is schematically reproduced as a background in Fig. \ref{fig:FeSC_OSM}).

Both the stoichiometric end members of the family have also been explored with isovalent chemical substitutions that act on the structure at fixed doping. In particular the substitution K $\rightarrow$ Rb, Cs acts as a negative chemical pressure and lengthen the Fe-Fe distance.

Electronic structure (typically density-functional theory - DFT) calculations show a complex of 5 conduction bands dispersing roughly W~$\sim$~4eV. These bands are mainly of character coming from all five Fe 3d orbitals, with some character of the ligand p-orbitals, and are populated by 6 electrons/Fe, in the stoichiometric parent compounds such as BaFe$_2$As$_2$. 
The same calculations predict the dominant magnetic orders (collinear antiferromagnetic in most of the compounds) and show, in the paramagnetic phase a semi-metallic bandstructure where the Fermi surface is made up of hole and electron pockets (respectively in the center and at the border of the Brilouin zone), as it is indeed verified in ARPES measurements. 
The nesting of such pockets is responsible, in the mainstream view\index{Itinerant electron scenario for Fe-based superconductors}, for the low-Temperature instabilities of the phase diagram, i.e. magnetism and superconductivity\cite{PaglioneGreene_NatRev}. 

This view based on itinerant electrons is indeed qualitatively quite successful, but substantial discrepancies between calculated and measured band structures and magnetic moments, plus the difficulties arising in explaining material trends as far as the superconducting properties are concerned point in the direction of electronic many-body correlations (basically neglected in DFT) playing a substantial role.

\begin{figure}[t!]
 \centering
 \includegraphics[width=0.6\textwidth]{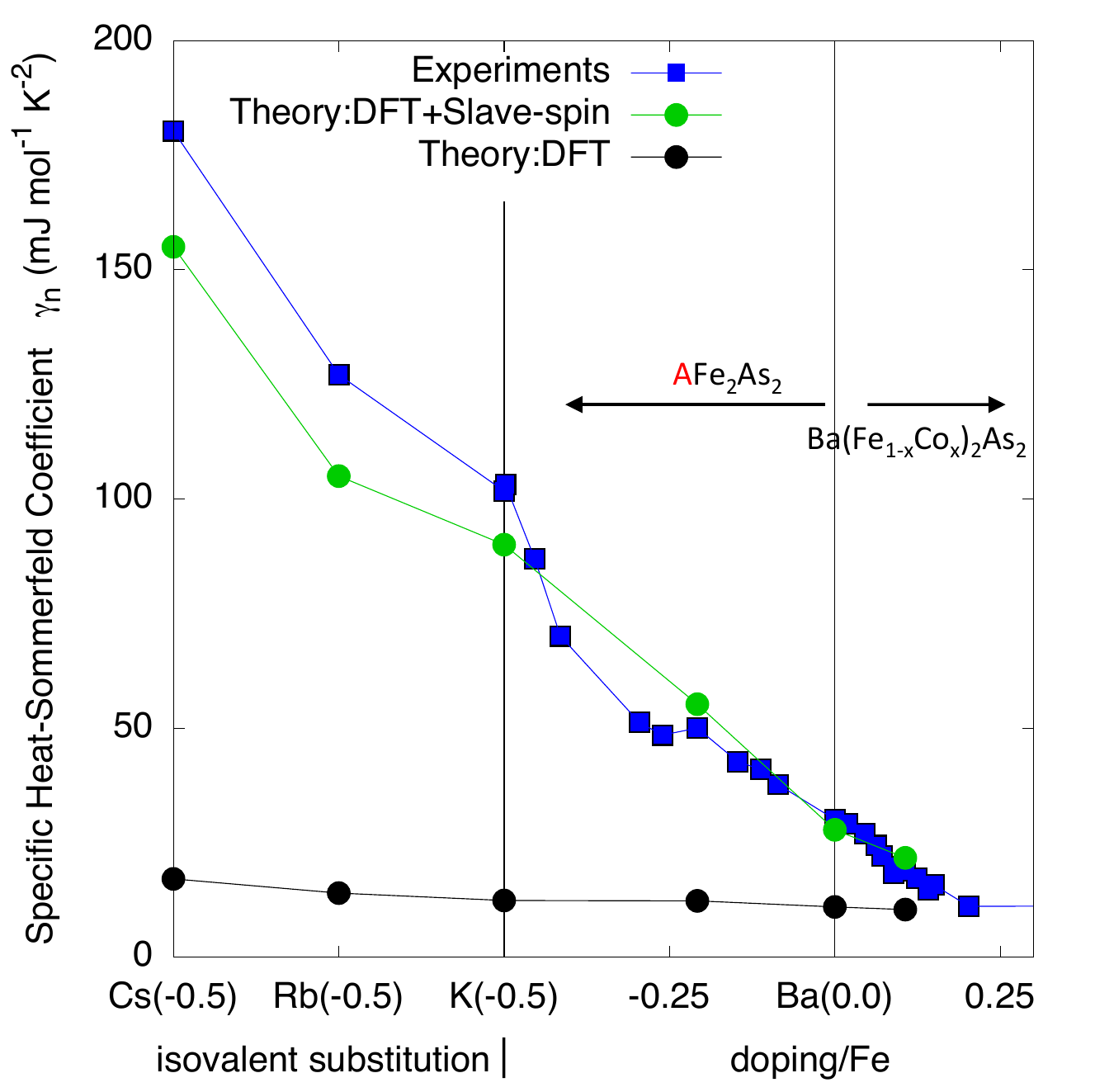}
 \caption{(Sommerfeld) coefficient of the linear-in-temperature contribution to the specific heat at low temperature in the normal phase in the122 family (under both electron and hole doping in BaFe$_2$As$_2$ and isovalent substitution in KFe$_2$As$_2$). The calculations including local electronic correlations (DFT+slave-spin mean field, green circles) are done with a unique choice of interaction parameters (U=2.7eV, J/U=0.25) for the whole family. The discrepancy between these calculations, that capture well the experimental data (blue squares), and those for uncorrelated electrons (black dots) shows that these compounds are indeed strongly correlated. Moreover, correlations increase monotonically with hole doping, supporting the fact that a Mott insulating state would be realized for half-filled conduction bands (doping of 1 hole/Fe). Isovalent substitution in KFe$_2$As$_2$ further strengthen this behaviour due to an expansion of the in-plane lattice parameters in the series K, Rb, Cs (adapted from Ref. \cite{Hardy_122_SlaveSpin_exp}, originally appearing in Ref. \cite{demedici_Vietri}).}
 \label{fig:Hardy_Sommerfeld}
\end{figure}

The most striking feature of these is quasiparticle mass renormalization.
This can be addressed by several experimental probes: low-T specific heat, optical conductivity, ARPES and quantum oscillations, to name the main ones. 
A collection of mass enhancement estimates by all these probes in the 122 family from the literature is shown in Fig. \ref{fig:FeSC_OSM}. Let's however focus first on the very fine measurement of the low-temperature slope (Sommerfeld coefficient)\index{Sommerfeld coefficient} of the temperature dependence of the specific heat in the normal phase, performed on the whole 122 family from Ref. \cite{Hardy_122_SlaveSpin_exp},\footnote{The extra points compared to the published plot, and relevant to the paramagnetic high-T phase in the zone of the phase diagram where the low-T phase is magnetic, appear already as unpublished material in Ref. \cite{demedici_Vietri} and are the result of a private communication with F. Hardy.} reported in Fig. \ref{fig:Hardy_Sommerfeld}.
This coefficient reads $\gamma_b=\pi^2k_B^2/3 N^*(\eps_F)$ and is directly proportional to the (quasi particle) density of states at the Fermi energy $N^*(\eps_F)$ \cite{Ashcroft}, which itself is enhanced for enhanced masses (i.e. for reduced dispersion of the bands - in a single band for instance one has $N^*(\eps_F)=(m^*/m_b) N(\eps_F)$, where $m^*$ and $m_b$ are the renormalized electron mass and the bare band mass, respectively).
From the figure it is clear (blue squares) that the Sommerfeld coefficient grows monotonically with diminishing electron density all across the family. The value reached for the end member KFe$_2$As$_2$ with density of 5.5 electrons/Fe is $\sim$100mJ/mol K$^2$, which approaches the ranges of heavy fermionic compounds. Remarkably, a further raise happens upon isovalent substitution, in the series K, Rb, Cs. These substitutions stretch the lattice parameters, thus reducing the bare electron hopping amplitudes, enhancing further the effect of electron-electron interactions. This is however a subleading effect, compared to the enhancement of correlations due to the doping. Indeed in the progressive substitution Ba $\rightarrow$ K there is actually a contraction of the Fe-Fe distance twice as large as the aforementioned stretch. Still the correlations increase enormously. Moreover a different hole-doping substitution, Fe $\rightarrow$ Cr, induces negligible changes in the Fe-Fe distance and has the same trend of steadily increasing correlations (see e.g. Ref. \cite{Lafuerza_XES_122} and references therein).

DFT calculations are unable to reproduce this trend. Indeed the calculated value of the Sommerfeld coefficient is of the order 10 mJ/mol K$^2$ or slightly above, throughout the family.

Dynamical many-body correlations can be included with several methods. On of the most popular is dynamical mean-field theory (DMFT). Here we report calculations within a method similar in spirit (i.e. it is a local mean field capturing the low-frequency part of a frequency-dependent self-energy), but much cheaper in terms of computational ressources, the Slave-Spin Mean Field (SSMF)\cite{demedici_Vietri}. \index{Slave-spin mean field} This is particularly suited to address the quasiparticle properties and their renormalizations in terms of the local interactions. The advantage of a simplified method is to be able to explore thoroughly the space of parameters (both compound-wise and interaction-wise) to highlight the main trends.

Details about this method can be found e.g. in Ref. \cite{demedici_Vietri}, let here just specify that it lumps the effect of the many-body local interactions into a renormalization of the hopping probabilities, in an orbitally resolved way. Indeed the starting point is a bare hamiltonian, typically a tight-binding fit of a DFT bandstructure parametrized by the hopping amplitudes $t^{lm}_{ij}$ (where $i$ and $j$ are the sites of the ionic lattice where the basis functions (typically localized Wannier-like functions) are centered, and $l$ and $m$ are the orbitals giving rise to the treated conduction bands), in which only the electrostatic effects of the electron-electron interactions are included. The dynamical part of these interactions (where we specify the intra-orbital interaction $U$, the inter-orbital interaction for anti-parallel spins $U-2J$ and the Hund's exchange energy $J$, further gained when electrons have parallel spins) treated in Slave-spins yields renormalization factors $Z_l<1$ that reduce the hopping amplitudes, leading to a new (quasiparticle) band structure analogously parametrized\footnote{This formula holds when $i\neq j$. Local orbital energies for $i=j$ are shifted by other effective orbital-dependent parameters $\lambda_l$.} by $\tilde t^{lm}_{ij}=\sqrt{Z_lZ_m}t^{lm}_{ij}$.

The Sommerfeld coefficient can thus be directly evaluated from the renormalized quasi particle density of states and as it can be seen in Fig. \ref{fig:Hardy_Sommerfeld} the DFT+SSMF calculations (obtained with a single set of interaction parameters, and varying only the ab-initio structure and the total electron density) capture the trend throughout the 122 family. This agreement with experiments together with the clear failure of the bare DFT calculations shows unambiguously the degree of electronic correlation of these compounds.

The same calculations explain easily the reason of this trend. Indeed the correlations increase monotonically with reducing the electron density throughout the range 6.2 to 5.5 electrons/Fe. The theoretical simulations show that when reaching 5 electrons/Fe ($d^5$ configuration) a Mott insulator\index{d$^5$ Mott insulator in Fe-based superconductors} is obtained. Thus in this picture electron masses increase until diverging for half-filled conduction bands. This Mott insulator, as we will see, is strongly favored by Hund's coupling and influences a large part of the phase diagram, even for densities as large as 1 electron/Fe away from it (the $d^6$ configuration of the parent FeSC) as it is clear from the above data, and beyond. 

Moreover BaMn$_2$As$_2$ (d$^5$) and BaCo$_2$As$_2$ (d$^7$) can also be synthesized and they are respectively a Mott insulator and an uncorrelated metal, clearly in line with the present picture.

Another recent experimental study\cite{Lafuerza_XES_122}\index{local moment in the normal phase of Fe-based superconductors} confirms this Mott-Hund scenario through the analysis of the formation of the local magnetic moment that fluctuates in the paramagnetic metal. Indeed the Mott insulating state is the loss of metallicity due to the local configurations becoming energetically very unfavorable to the charge fluctuation necessary for electrons in a metal to flow. In a Mott state where Hund'coupling dominates this reduction of charge fluctuations together with the tendency of electron spins to align favors the highest possible atomic spin configurations. Thus naturally upon approaching the half-filled Mott-Hund insulating state we expect to see these high-spin configurations gradually prevailing and building up a large local moment.

\begin{figure}[t!]
 \centering
 \includegraphics[width=0.6\textwidth]{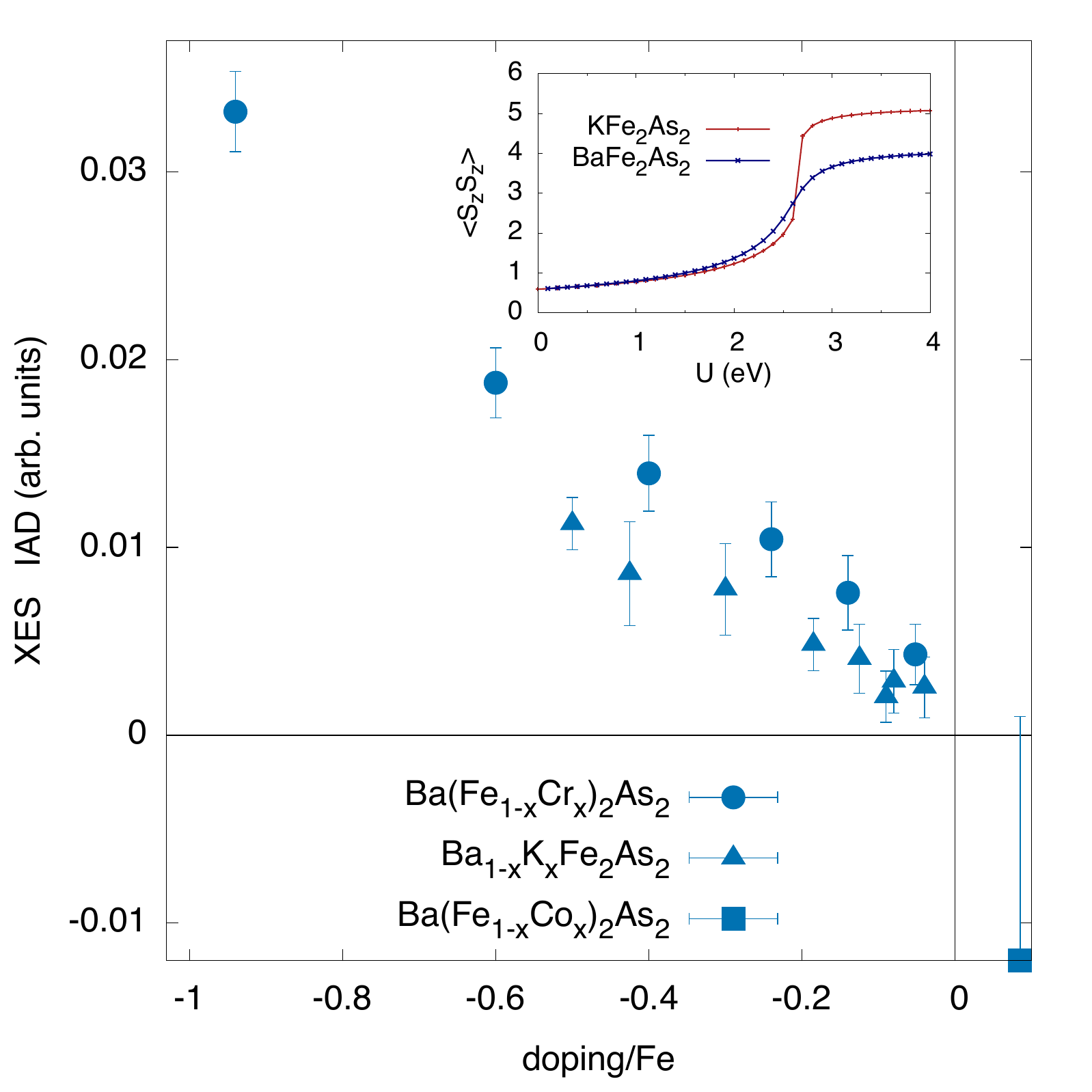}
 \caption{Evolution of the Fe 3d local magnetic moment in the paramagnetic normal phase for doped BaFe$_2$As$_2$ estimated with X-ray Emission Spectroscopy at room-temperature. IAD is the result of a deconvolution method, yielding a value that scales with the Fe 3d local moment\cite{Gretarsson_LocalMoments,Lafuerza_XES_122}. The local moment increases monotonically with hole doping (which is obtained with two different chemical substitutions yielding very similar results) and in comparison with the electron doped material, in line with the Mott-Hund scenario. Inset: DFT+Slave-Spin estimate of the local moment in BaFe$_2$As$_2$ and KFe$_2$As$_2$ as a function of the interaction strength U: a considerable increase is realized only for U large enough to be in the Hund's metal regime. In the DFT limit (U=0) the moment is basically unchanged between the two compounds.(adapted from Refs. \cite{Lafuerza_XES_122} and \cite{Gretarsson_LocalMoments})}
 \label{fig:Lafuerza_XES}
\end{figure}

In a non-magnetic phase these local moments do not form a static long range order and thus a fast spectroscopic technique sensitive to the size and not the direction of these fluctuating moments is needed to characterize them. X-ray emission spectroscopy (XES)\index{X-ray Emission Spectronscopy and local moments} is such a technique. Indeed it probes the energy of a photon emitted in a deexcitation from a valence state (in the present case the Fe 3p) into a core hole previously created by incident radiation. The decaying electron can have spin up or down and its energy is different in these two configurations when a net magnetic moment is present in the near Fe 3d open shell. Thus the spectroscopical line due to the photons emitted in the deexcitation splits, in a proportional way to the magnitude of the local moment, which can be thus characterized\footnote{Absolute measurements of the moment are made uneasy by the line shapes and intensities and by the proportionality factor which is a screened exchange constant not readily obtained (see and \cite{Gretarsson_LocalMoments,Lafuerza_XES_122} and references therein). This constant is however believed to vary slowly within the series of doped BaFe$_2$As$_2$ reported here.}. 
A measure of this splitting (the so called IAD value\cite{Gretarsson_LocalMoments,Lafuerza_XES_122}) for hole-doped BaFe$_2$As$_2$ series is reported in Fig. \ref{fig:Lafuerza_XES} and compared with that of the electron doped compound. Indeed a monotonic increase of the Fe 3d local moment is observed with reducing density, throughout the phase diagram.

Calculations within DFT+SSMF (inset of Fig.\ref{fig:Lafuerza_XES}) again show that this is a clear indication of FeSC being in a zone of influence of the half-filled Mott insulating state. Indeed the estimate of the local moment from this method shows, both in BaFe$_2$As$_2$ (d$^6$ configuration) and in KFe$_2$As$_2$ (d$^{5.5}$ configuration), a clear crossover as a function of the interaction strength U (at fixed $J/U$) between a low-moment (at small U) and a high-moment (at large U) region. The frontier departs from the critical U for the Mott transition at half-filling and moves very slightly to larger values with doping, the crossover becoming progressively smoother. From the data it is clear that an increase in the local moment in going from BaFe$_2$As$_2$ to KFe$_2$As$_2$ as seen in the experiments is only possible if the materials are in the large-moment zone.

The third experimental feature well captured by realistic theoretical calculations we want to highlight is the orbital selectivity of the electron correlation strength. 
 
This is illustrated, across the 122 family in the upper panel of Fig. \ref{fig:FeSC_OSM}. Indeed several experimental probes are sensitive to the mass enhancement and by collecting the various estimates (from specific heat, optical conductivity, ARPES and quantum oscillations and their comparison to DFT estimates - details are given in Ref. \cite{deMedici_OSM_FeSC}) one finds a substantial agreement on values between 2 and 3 for the electron-doped side, and an increasing disagreement with hole doping. 

This can be clearly interpreted in terms of increasingly differentiated correlation strengths leading, at strong hole doping, to the coexistence between strongly correlated and weakly correlated electrons. These electrons enter differently in the the different experimental quantities and this explains the diverging estimates. For example the increase of the Sommerfeld coefficient $\gamma$ we have already seen in this section is caused by an increase of the density of states, which is a sum over the orbital index $l$ of contributions $\sim (m^*/m_b)_l$. The estimates of correlations in optical conductivity instead is done comparing the measured low-frequency (Drude) spectral weight $D$ with the theoretical one calculated in DFT (which in a single band case is inversely proportional to the band mass). In the multi-band case the renormalized Drude weight is then a sum of terms $\sim (m_b/m^*)_l$. If the mass renormalization were the same for all electrons then the common renormalization factor would factorize and two estimates of correlation strength $\gamma(\mbox{measured})/\gamma_{DFT}$ and $(D(\mbox{measured})/D_{DFT})^{-1}$ would coincide.
Instead if the various $(m^*/m_b)_l$ differ, as it happens for series or parallels of resistances, $\gamma$ will be dominated by the largest mass enhancement, and $D$ by the smallest.

Analogously, the renormalization of the whole bandstructure (labeled "ARPES whole" in Fig. \ref{fig:FeSC_OSM}), this being an entangled complex where all orbital contributions mix, will be renormalized as the least renormalized of the contributions.

On the other hand ARPES can measure the renormalization in a band/orbital resolved fashion, by comparing the slope of the dispersion (the Fermi velocity, $v_F^*=v_F(m_b/m^*)$ for a single band model) measured in different points of the Fermi surface of different orbital character to that coming from band structure calculations. The estimates from these different measures are reported as orange crosses in Fig. \ref{fig:FeSC_OSM} and it is clear that they spread increasingly with hole-doping, confirming the increasing differentiation of the correlation strength. This is further confirmed by analogous band-resolved estimates from quantum oscillations (yellow crosses).

This differentiation unambiguously comes out of the same theoretical calculations that reproduce the values of the Sommerfeld coefficient in Fig. \ref{fig:Hardy_Sommerfeld} and the increase of the local moment in Fig. \ref{fig:Lafuerza_XES}. Indeed in the lower panel of Fig. \ref{fig:FeSC_OSM} the orbitally-resolved renormalization factors $(m^*/m_b)_l=Z_l^{-1}$ calculated in DFT+SSMF for doped BaFe$_2$As$_2$ (lines) and for KFe$_2$As$_2$ (squares)\footnote{Calculations are done for a given DFT structure. The lines are obtained varying the electron filling for the DFT structure of BaFe$_2$As$_2$. The squares are values calculated for the KFe$_2$As$_2$ structure.} are shown. Clearly there is a crossover, roughly around the electron density of the parent compound, between a region at electron doping in which the renormalization is similar for all the electrons, and another at hole doping where the differentiation is strong. The differentiation increases the closer the density to half-filling, where the Mott insulating state is obtained.

Summarizing, the three features mentioned in the introduction, namely: i) strong correlations due to a Mott insulating state realized for the half-filled system and increasing when approaching this filling, ii) a correspondingly increasing local moment in the metallic paramagnetic phase, and iii) a strong differentiation of the correlation strength among electrons of different orbital character, are all realized and clearly seen experimentally in the 122 family of FeSC. They happen after a crossover\index{Hund's metal crossover, in Fe-based superconductors} roughly located around the filling of the parent compound BaFe$_2$As$_2$, so that while the hole-doped side clearly shows these features, the electron-doped side has them much less pronounced and fading into a more common uncorrelated metallic phase. 

Calculations within DFT+SSMF correctly describe all this physics within an unbiased unique choice of interaction parameters for the whole family, highlighting the paramount role played by the filling of the conduction bands, and how the distance from the half-filled Hund's induced Mott insulator is a key quantity dominating the many-body physics in these compounds. The same calculations show that the crossover between the normal and the Hund's metal is a frontier in the U/density plane departing from the Mott transition and moving slightly to higher U values with doping, such that it can be crossed both acting on the doping and on the interaction strength U.

We take these features as defining the Hund's metal, so that it appears, based on the above analysis, that the parent compound with density of 6 electrons/Fe of the 122 FeSC family are located in the proximity of the crossover. Isovalent doping\cite{PaglioneGreene_NatRev} can be performed on BaFe$_2$As$_2$ too with the substitution As $\rightarrow$ P which reduces the in-plane Fe-Fe distance. Albeit rich of further phenomenology that we will not detail here evidences are that this substitution, acting as positive pressure (equivalent to reducing U in our calculations), brings the metal towards a weakly correlated phase, as again predicted by DFT+SSMF calculations.

\section{Model studies: generality of Hund's metals main features}\label{sec:models}

The goal of this section is to make a parallel analysis to the one performed in Sec. \ref{sec:abinitio-exp} on the main features of the Hund's metal which are apparent in the experiments on FeSC and well captured by DFT+Slave-spin calculations.
The aim is to show that these features appear identically in models with featureless bare hamiltonians, in the proximity of a Hund's-induced half-filled Mott insulator, and are thus the outcome of the local many-body physics dominated by Hund's coupling and relatively independent from the Fermiology and the specificities of the band structures. 

Indeed let's focus on the multi-orbital Hubbard model of Hamiltonian:
\be\label{eq:multiorb_Hubbard}
\hat H = \sum_{i\neq j lm\s} t^{l m}_{ij} d^\dag_{il\s}d_{jm\s}+\sum_{il\s}(\eps_l-\mu)n^d_{il\s}+\hat {H}_{int}
\ee

where $d^\dag_{il\s}$ creates an electron in orbital $l$ with spin $\s$ on site $i$ and we take $t^{l m}_{ij}=\delta_{lm}t$ equal for all bands and all orbitals degenerate $\eps_l=0, \; \forall l$ (the total filling being tuned by the chemical potential $\mu$). The lattice geometry is such to have a semicircular bare density of states (DOS) for each band (Bethe lattice) of half-width D=2t.
The interaction Hamiltonian reads\footnote{The interaction Hamiltonian eq. (\ref{eq:H_int}) is an approximation of the more rigorous "Kanamori" hamiltonian in which, besides the density-density terms here considered, off-diagonal "spin-flip" and "pair-hopping" terms are present. The approximation of dropping them is however quite customary for computational reasons.}:
\be
\hat H_{int}\,=\,U\sum_l n_{l\up}n_{l\down}\,+\,U^\prime\sum_{l\neq m} n_{l\up}n_{m\down}\,
+(U^\prime-J) \sum_{l<m,\sigma} n_{l\s}n_{m\s},
\label{eq:H_int}
\ee
where we make the with typical choice $U'=U-2J$ (a discussion on this prescription can be found in \cite{georges_annrev}) and the three contributions mentioned in the previous section are easily read out\footnote{This is the same Hamiltonian form that is used in the DFT+SSMF simulations discussed in the previous section. The only difference is in the choice of the $t^{l m}_{ij}$ which are fitted on a DFT bandstructure, yielding a 5-orbital model in the case of typical FeSC}.

We solve this model within the slave-spin mean field\cite{demedici_Vietri}. 

The important point we want to highlight here is that the main features of the Hund's metal are present already in a simple 2-orbital Hubbard model in presence of Hund's coupling. 

\begin{figure}[t!]
 \centering
 \includegraphics[width=0.8\textwidth]{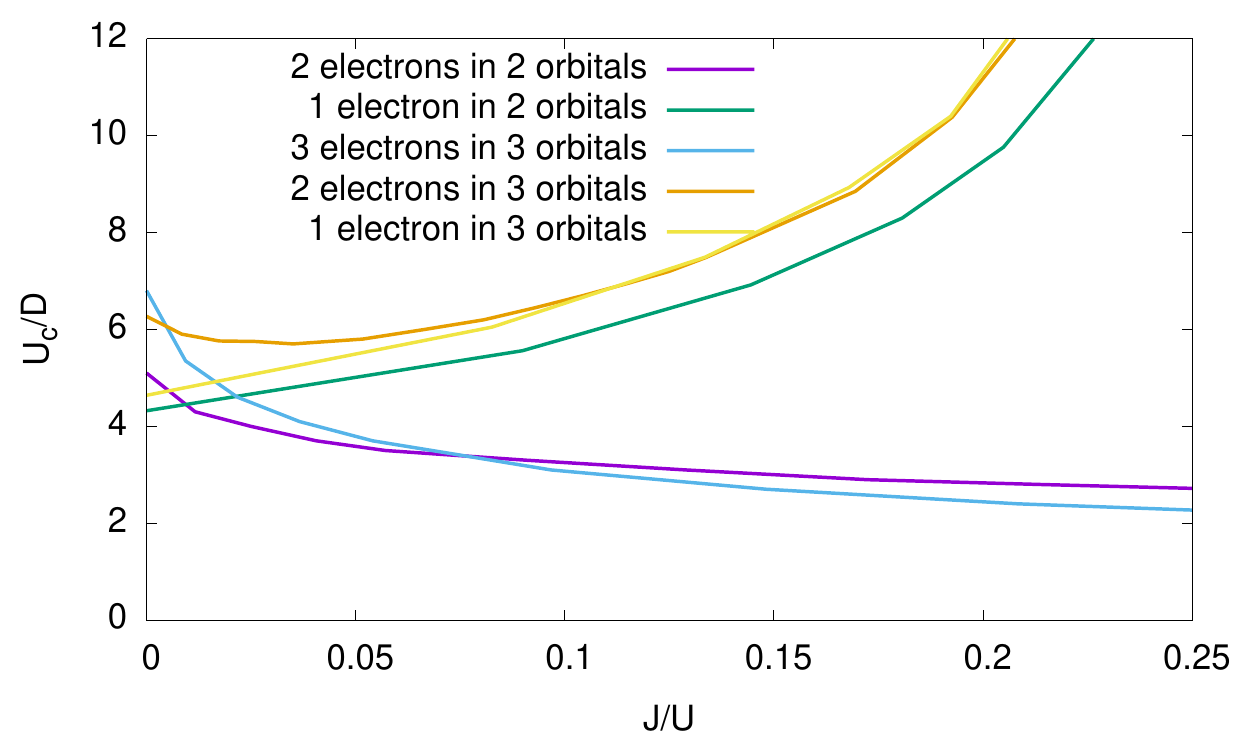}
  \caption{Critical interaction strength $U_c$ for the Mott transition in the 2-orbital and 3-orbital Hubbard models with semi-circular DOS of bandwidth $W=2D$, for all distinct integer fillings, as a function of Hund's coupling $J/U$ (adapted from Ref. \cite{demedici_MottHund}).}
 \label{fig:Uc_Hund}
\end{figure}

Indeed as seen in the previous section Hund's coupling favors a Mott insulator at half-filling (2 electrons in 2 orbitals, in this case), while the Mott transitions for the other possible integer fillings (1 and 3 electrons in 2 orbitals - which are physically identical due to the particle-hole symmetry of the semi-circular DOS) are sent to very high interaction strength.
In Fig. \ref{fig:Uc_Hund} we report the critical interaction strength for the Mott transition calculated within SSMF\footnote{These calculations were done including spin-flip and pair-hopping terms, but are very close (and qualitatively identical) to the result of the present model in which we neglect these terms} and indeed it is clear that for the customary value\footnote{The typical ab-initio estimates for 3d transition metal compounds is rather $J/U\simeq 0.12\div 0.15$, however it was shown\cite{demedici_Vietri} that in the present approximation $J/U=0.25$ is a suitable choice to reproduce DMFT results with spin-flip and pair hopping at $J/U\simeq 0.15$, in the typical fillings of interest not far from half.} $J/U=0.25$ the Mott transition at half-filling is brought to much lower values of U compared to the $J=0$ case, while the opposite happens at the other integer fillings.

Identical trends are found for Hubbard models with 3-orbitals\cite{georges_annrev} or more\cite{demedici_Vietri}: at half-filling the critical interaction strength $U_c$ in presence of sizable Hund's coupling (say $J/U=0.25$) is a fraction of the bandwidth, whereas for all other fillings it is several times the bandwidth.

\begin{figure}[t!]
 \centering
 \includegraphics[width=0.4\textwidth]{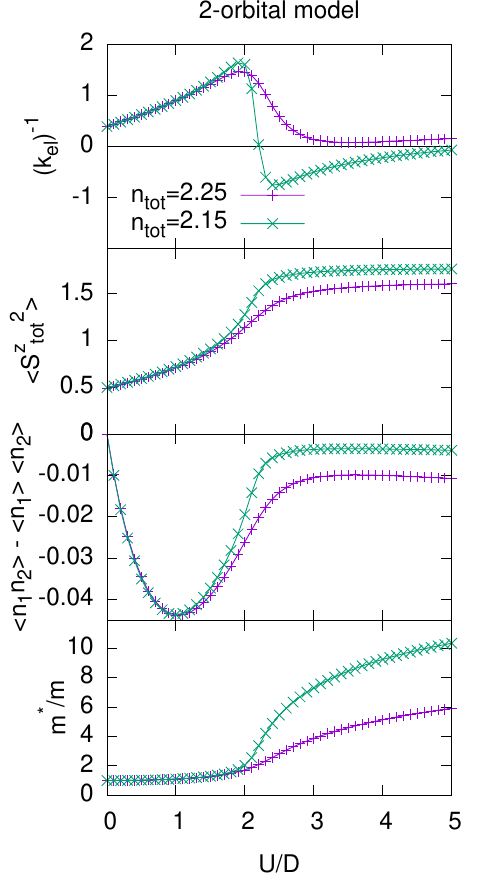}
 \includegraphics[width=0.4\textwidth]{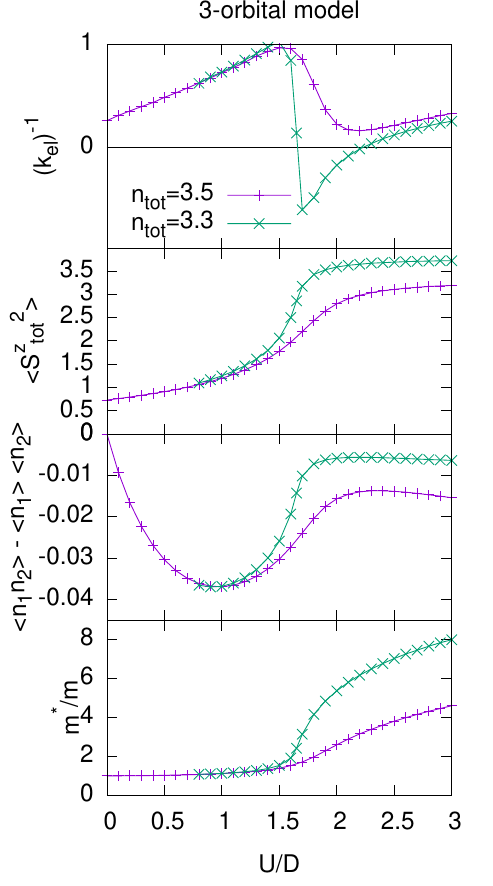}
  \caption{Main quantities highlighting the crossover from normal (small U/D) to Hund's metal (large U/D) in 2-orbital (left) and 3-orbital (right) Hubbard model with Hund's coupling J/U=0.25, solved within slave-spin mean-field for several dopings in proximity of half-filling: (inverse) electronic compressibility, total local moment, inter-orbital charge correlation function, mass enhancement (from Ref. \cite{demedici_compress}). The analogy with the analogous quantities calculated in realistic simulations validated by experiments shows the robustness of the Hund's metal physics by respect to material details.}
 \label{fig:crossover}
\end{figure}

Moreover the half-filled Mott insulator dominates the phase diagram, for $U\gtrsim U_c$ for an extended range of filling even quite far from half. 
As an example we report in Fig. \ref{fig:crossover} several quantities calculated at fixed density that undergo a crossover\index{Hund's metal crossover, in models} as a function of U on a frontier departing from the $U_c$ at half-filling. These quantities are the mass enhancement (bottom panel), the inter-orbital charge correlations $\langle n_1n_2\rangle-\langle n_1\rangle \langle n_2\rangle$ (where $n_l=\sum_\s n^d_{il\s}$, lower middle) and the local moment (upper middle). The top panel reports the inverse of the electronic compressibility that we will discuss later in the chapter.

\begin{figure}[t!]
 \centering
  \includegraphics[width=0.4\textwidth]{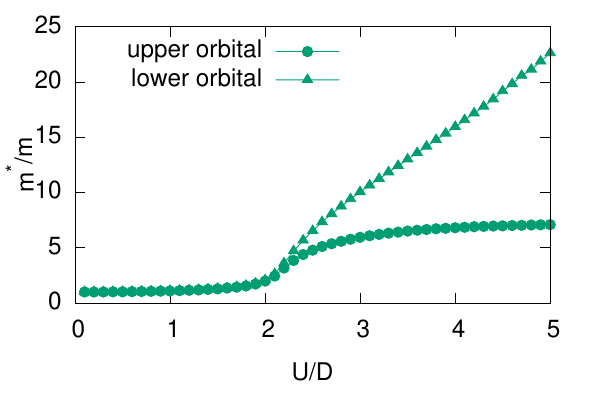}
  \caption{As in the left panel of Fig. \ref{fig:crossover} (2-orbital Hubbard model, J/U=0.25) with total density $n_{tot}=2.15$, but with a small crystal field splitting $\eps_1-\eps_2=0.05D$, in order to show the tendency to orbital selectivity upon entering the large U region beyond the Hund's metal crossover.}
 \label{fig:orbsel}
\end{figure}

Obviously the mass enhancement and the local moment undergo the same kind of crossover described in the previous section. The mass enhancement clearly goes from a hardly changing value near 1 at low U to a much larger value increasing with U. 
Depending on the proximity to half-filling ($n_{tot}=2.0$) the effect is more or less pronounced.
The local moment undergoes a rapid change of behaviour from a low value increasing from the uncorrelated one at small U to a saturated high-value at large U, here again the closer to the saturated value for the Mott insulator the nearer the density is to half-filling.
The analogy between the local moment behaviour in the model plotted in the upper-middle panels in Fig. \ref{fig:crossover} and the behaviour calculated in the ab-initio simulations for BaFe$_2$As$_2$ and KFe$_2$As$_2$ in Fig. \ref{fig:Lafuerza_XES} is obvious.

Analyzing the the third feature, orbital selectivity, is more subtle here, since the model is perfectly degenerate between orbitals 1 and 2, and thus the mass enhancement will be the same for the two orbitals, by symmetry.
The tendency towards orbital selective correlations can be however highlighted by looking at the inter-orbital charge-charge correlations.
This correlation function is obviously zero in the uncorrelated limit U=0 and it grows initially with the interaction strength. However in proximity of the crossover it undergoes a quick suppression, testifying the independence of the charge fluctuations between the two orbitals in the large U phase. Charge excitations, which are ultimately responsible for the metallicity and the suppression of which leads towards the Mott insulating state, being independent in each orbital allow for a different individual proximity of each orbital to the Mott state. This has been termed "orbital decoupling"\cite{demedici_MottHund,georges_annrev,deMedici_OSM_FeSC,demedici-SpringerBook,demedici_Vietri}\index{orbital decoupling}.
Indeed the same correlation function is reported for BaFe$_2$As$_2$ in Refs. \cite{deMedici_OSM_FeSC,demedici_Vietri} (obviously in a realistic case all pair of orbitals will give rise to a different correlation function) and the behavior over the whole phase diagram is completely analogous.

A more direct check of the link between the orbitally-decoupled charge excitations and the orbital-selectivity of the correlation strength in the present model case can be obtained introducing a small crystal field splitting, e.g. $\eps_1-\eps_2=0.05D$, i.e. 1/40th of the bandwidth.
As can be seen from Fig. \ref{fig:orbsel} this results in a clear orbital selectivity of the mass enhancements, starting at the crossover and growing quickly with U. The orbital closer to individual half-filling (orbital 2, which is lowest in energy for a total filling $n_{tot}$=1.85) is the most correlated of the two, following the orbital-decoupling physics.

All the same crossovers can be observed with similar doping-dependent plots at constant U. Moreover a completely analogous behavior is found in the 3-orbital Hubbard model (right panels in Fig. \ref{fig:crossover}) and for a larger number of orbitals\cite{demedici-SpringerBook}).

Thus in conclusion we have shown that the three features that we have taken as a definition of a Hund's metal and that are found in realistic simulations of FeSC and confirmed by experiments, are also identically found in degenerate models with featureless semi-circular densities of states. 
Irrespectively to the number of orbitals a Mott insulator is favored at half-filling by Hund's coupling (and is found for $U_c$ of order the bandwidth or smaller for realistic Hund's coupling) and dominates a large range of the U-doping phase diagram for U$\gtrsim U_c$.
Identically too, the crossover into the Hund's metal phase happens on a frontier stemming from the Mott transition point at half-filling and extending at finite doping for a large range of dopings even far from half-filling.

This robustness is due to the local many body physics being the cause of these distinctive features, and to the fact that the details of the bare band structure enter through few local parameters: crystal-field splitting, kinetic energy (i.e. the first moment of the density of states), possibly orbitally resolved, etc. This is the reason why such a phenomenology can be common to many different materials irrespectively of the details of the band structure (and of the Fermi surface most notably), and can be righteously labeled as a general behavior, the Hund's metal behavior.

\section{Analytical insight into Hund's metal mechanisms}\label{sec:analytics}

In this Section we will give analytical arguments that provide some insight in the Hund's metal phenomenology outlined thus far.
These arguments are based on an analysis of the spectrum of the half-filled Mott insulator that was seen to influence a large zone of the U-doping parameter space in the previous sections, and is here taken as responsible for the Hund's metal phenomenology.

The spectrum of a Mott insulator can be analyzed in terms of the excitations of the system in the atomic limit, i.e. with all hoppings in eq. (\ref{eq:multiorb_Hubbard}) $t^{l m}_{ij}=0$. Indeed for the 2-orbital model with $\eps_1=\eps_2=0$ and $\mu=(3U-5J)/2$ which ensure hal-filling for a particle-hole symmetric DOS\cite{demedici_Vietri} the spectrum is (the zero of energy is arbitrarily fixed at the ground state energy):

\be\label{eq:spectrum}
\left\{
\begin{array}{cc}
|\up\down,\up\down\rangle |0,0\rangle & E=2U-2J\\
&\\
\begin{array}{c}
 |\up\down,\up\rangle |\up\down,\down\rangle |0,\up\rangle |0,\down\rangle \\ 
|\up,\up\down\rangle |\down,\up\down\rangle |\up,0\rangle |\down,0\rangle 
\end{array}
& E=\frac{U+J}{2}\\
&\\
\begin{array}{c}
|\up\down,0\rangle \qquad |0,\up\down\rangle
\end{array}
& E=3J\\
\begin{array}{c}
|\up,\down\rangle \qquad |\down,\up\rangle  
\end{array}
& E=J\\
\begin{array}{c}
|\up,\up\rangle  \qquad |\down,\down\rangle \\
\end{array}
& E=0\\
\end{array}
\right.
\ee

The Coulomb repulsion U splits the sectors with the same total charge while J splits the half-filled sector depending on the spin alignment of the two electrons and of them paying inter- or intra- orbital repulsion. As a result the ground state is 6 times degenerate at J=0, while the high-spin doublet is selected for nonzero J\footnote{The present discussion, besides a modification of the spectrum that has no impact, is equally valid for the Kanamori hamiltonian for which the ground state is a high-spin triplet.}.

The ground state of the lattice system in the atomic limit will then be that in which every site hosts two electrons in one of the high-spin configurations. 
The gap to overcome to establish conduction in such a system amounts to the energy needed to move an electron from one site to another
and is thus the sum of the energies necessary for adding a particle on an atom and for subtracting one on another , that is: 
\be
[E(n+1)-E(n)]+[E(n-1)-E(n)]= E(n+1)+E(n-1)-2E(n),
\ee
where $E(n)$ is the atomic ground state with n particles. It can be directly read from the above scheme, where both energy differences read $(U+J)/2$ so that spectral function has two delta-like features at $\pm(U+J)/2$, and the total atomic gap reads $\D_{at}=U+J$.

Upon reintroduction of the hopping these features in the spectrum broaden in two "Hubbard" bands due to the delocalization of the charge excitations. Indeed at zero hopping there are many degenerate excited states, one for each lattice site, since an extra electron (or an extra hole) with a given orbital and spin flavor can be added on any site. The hopping connects them and removes the degeneracy, spreading the states over a range roughly the bandwidth W, in analogy with the case of a non-interacting electron.

\begin{figure}[t!]
 \centering
  \includegraphics[width=0.6\textwidth]{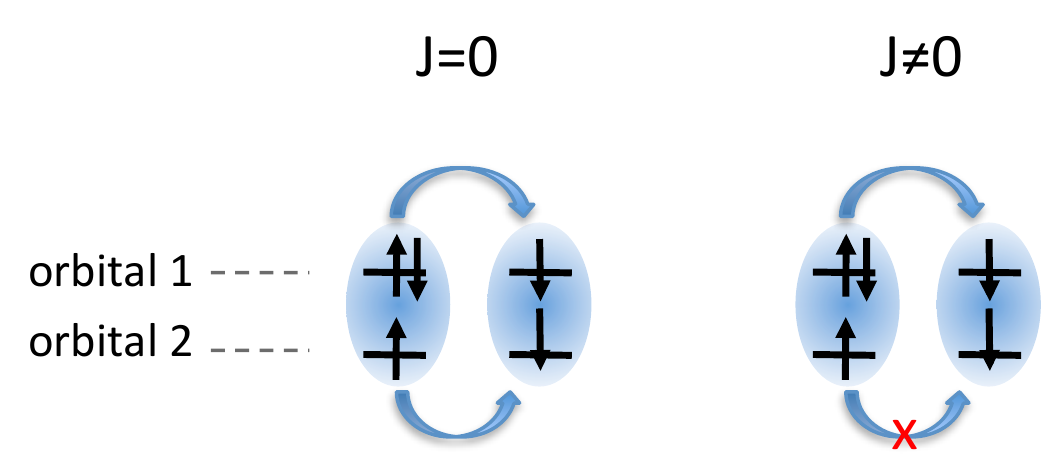}
  \caption{Delocalization of charge excitations. In absence of Hund's coupling ($J=0$) all 6 configurations with 2 electrons on a given site are degenerate (see eq. (\ref{eq:spectrum})). Thus both hopping channels between two sites (light blue arrows) are allowed without an extra energy cost. For finite J instead the only hopping allowed is in the channel where the extra electron was created, since hopping in the other channel would produce a two-electron configuration with a doubly occupied orbital, which has now a higher energy.}
 \label{fig:Hoppings}
\end{figure}

It should be noticed that at J=0 in a multi-orbital model (take for simplicity only diagonal hopping in the orbital index) the spread is actually larger\cite{gunnarsson_fullerenes}, of order $\sim \sqrt{M}W$, where M is the number of orbitals. Indeed (see Fig. \ref{fig:Hoppings}) from the site where the extra electron is created, say in orbital 1 - and that thus hosts 3 electrons - not only hopping from orbital 1 but also hopping from orbital 2 connects this state with another state of the same energy. Indeed this second hopping process leaves behind a site with a doubly occupied orbital and the other empty, which at J=0 is degenerate with all the other configurations of two electrons on a site. 

However when J is nonzero this extra degeneracy is removed. Indeed the two-electron configuration with orbital 1 doubly occupied left behind by the second hopping process at finite J is no longer degenerate with the high-spin configurations of the atomic ground state.
Thus an extra electron created in orbital 1 can only delocalize through the hopping process in its own channel, i.e. the charge excitation cannot take advantage of the multi-orbital nature of the system to delocalize and the width of the Hubbard bands becomes of order $\sim W$ again. This shrinking was also verified within dynamical mean-field theory, where the spectral function for a Mott insulator can be directly calculated and the width of the Hubbard bands measured\cite{demedici_Vietri}.\index{width of the Hubbard bands}

All in all this means that the gap between the two Hubbard bands, which are $\D_{at}$ apart is $\D\simeq \D_{at}-W=U+J-W$.
This gap will then close at interaction strength $U_c\simeq W-J$, which is of size of order of the bandwidth or less, and decreasing with increasing J, perfectly in line with what is found numerically\cite{demedici_Vietri}.\index{half-filled Mott insulator favored by Hund's coupling}

This argument can be generalized to any number of orbitals and explains why the half-filled Mott insulating state is favored by Hund's coupling. Indeed the half-filled sector is the one with a larger number of spare spins to align in order to gain exchange energy. Its distance in energy from all other sectors will grow with J and so will the Mott gap, needing thus a smaller U to close.

The same kind of arguments can be applied to the Mott transition at other integer fillings\cite{demedici_MottHund}, where however the effect is opposite. For instance in the 2-orbital model for filling of 1 electrons/site, the atomic ground state will be in the n=1 sector. This sector is unaffected by J, while the excited state with n+1 particles will be the ground state of the atomic n=2 sector. The energy of this state lowers with J, and thus J helps closing the gap in this case, thus disfavoring the Mott transition.

The general outcome\cite{demedici_MottHund} is that the critical interaction strength for the Mott transition in a system with M orbitals at large J goes like:
\be
U_c(n)\propto\begin{cases}
    3J, & \; \forall n\neq N \quad \text{(off half-filling)},\\
    -(M-1)J,  & \; n=M  \quad \text{(half-filling)}.
  \end{cases}\label{eq:D_at_J}
 \ee
which explains why for values of $U\sim W$ as it happens in FeSC and in the related models with sizable J/U, the Mott insulating state is only realized at half-filling.

This analytic arguments justifies the first two features of the Hund's metals. 
Indeed the electronic correlation strength naturally grows with reducing doping from the Mott insulator.
Also the local moment is maximized in the half-filled Mott insulator, in which the ground state lies in the sector with the highest possible spin configuration and charge fluctuations are minimized. Upon doping this sector will mix increasingly with the other charge sectors, where lower spin configuration are realized, and the resulting local moment will gradually decrease.

The argument above on J decoupling the hopping channels for the charge excitations in the various orbitals is a support to the third Hund's metal feature we have outlined, i.e. the role of Hund's coupling as an orbital decoupler\index{orbital decoupling} in general, favoring orbital-selectivity in the proximity of the half-filled Mott insulator.
Indeed in a system where the two orbitals differ, be it for the energy $\eps_1\neq\eps_2$ or for the hopping integrals (or both), the Hubbard bands in the spectral function will differ for the two orbitals\cite{Koga_OSMT}. This implies different gaps and can lead to orbitally-selective Mott transitions, if U is such that the gap is open only for one of the orbitals, and closed for the other. Analogously in the doped case the chemical potential can fall in the gap for one orbital and in the Hubbard band for the other causing again selective localization. In a more realistic case where off-diagonal hoppings are present and thus the character of the orbital mixes, one can expect that orbital selective Mott phases turn into metallic phases with different correlation strength, and this is indeed what is observed in simulations.

These arguments for independent gaps are just indicative for the electron correlations in the metallic phase. There indeed, more rigorous arguments for the low-energy long lived quasiparticle excitations should be used. SSMF offer a framework where this can be done and a low-energy analysis supporting the role of Hund's coupling as an orbital decoupler was done in Ref. \cite{demedici_Vietri}. 

\section{Compressibility enhancement and quasiparticle interactions}\label{sec:compress}\index{electronic compressibility}

In this last section it is worth mentioning some more recent work highlighting another feature connected with the normal to Hund's metal crossover. This feature emerges clearly from theoretical calculations: an enhancement (culminating in a divergence) of the electronic compressibility\index{electronic compressibility} near the crossover.

\begin{figure}[t!]
 \centering
 \includegraphics[width=0.4\textwidth]{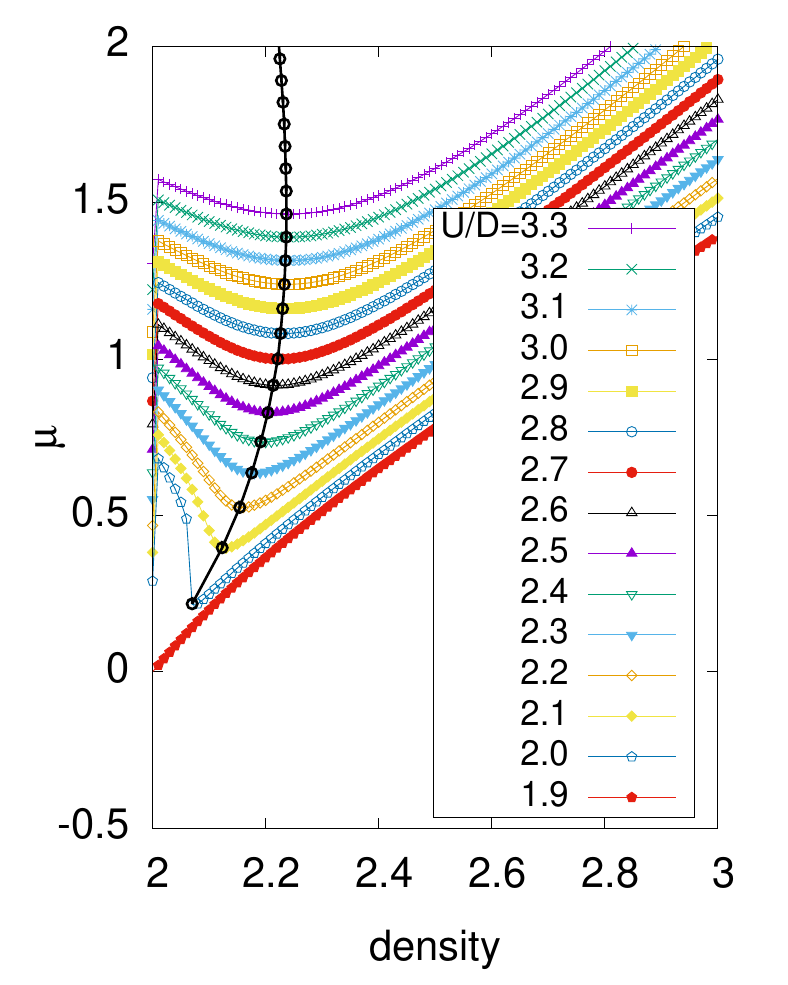}
 \includegraphics[width=0.4\textwidth]{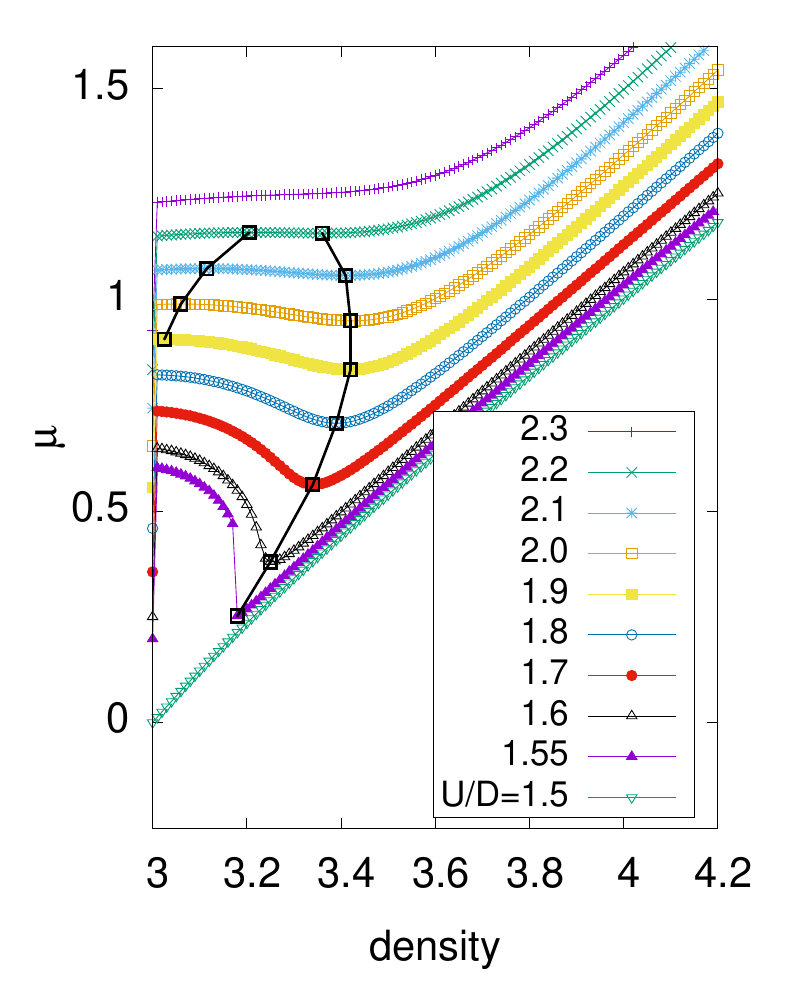}
 \includegraphics[width=0.5\textwidth]{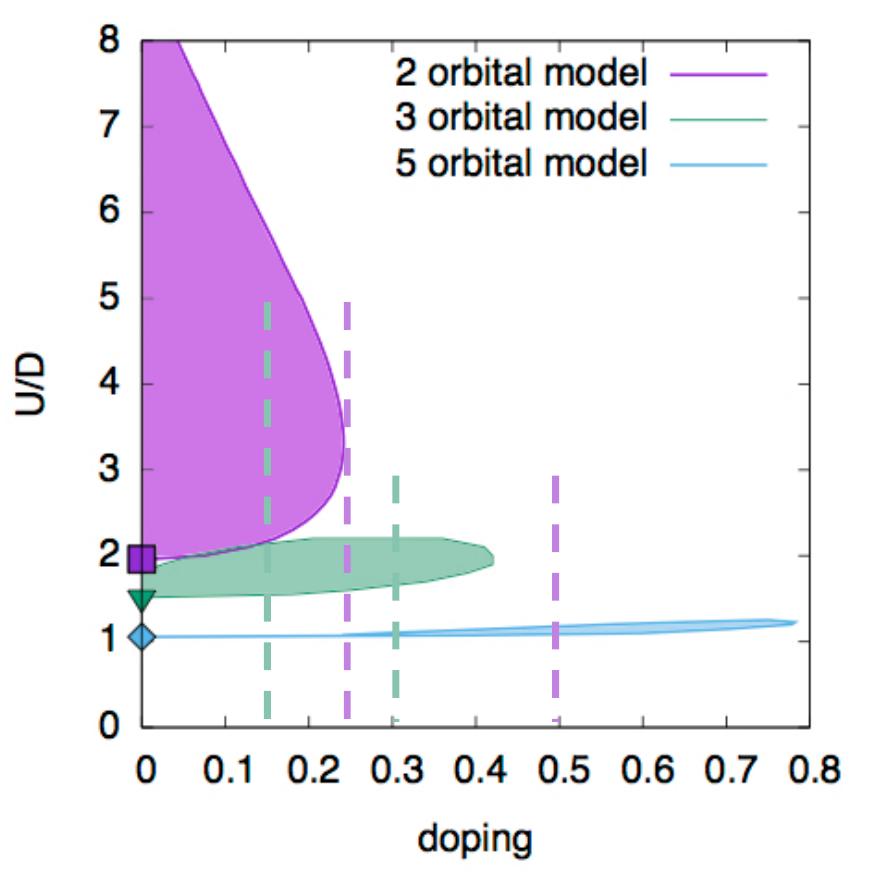}
  \caption{Upper panels: chemical potential vs density curves for the 2-orbital (left) and 3-orbital (right) Hubbard models with $J/U=0.25$ calculated within DFT+SSMF. For $U>U_c$ (the critical interaction value for the Mott transition at half-filling, with $U_c/D=1.96$ in the 2-orbital model and $U_c/D=1.515$ in the 3-orbital model) the curves show a negative slope, indicating an unstable system for a range of doping inside a spinodal line (black dots) which is the locus where the compressibility $\kappa_{el}={dn}/{d\mu}$ diverges. In the 3-orbital model (and in the 5-orbital model - not shown) a second spinodal line where the compressibility becomes positive again delimitates a region in the shape of a "moustache". The instability regions for all these models are reported (colored areas) in the lower panel, in the interaction-density plane. The dashed lines represent the scans in the space of parameters corresponding to the calculations reported in Fig.~\ref{fig:crossover}  (from Ref. \cite{demedici_compress}).}
 \label{fig:compress}
\end{figure}

Indeed when investigating within SSMF the proximity of the half-filled Mott insulator in presence of Hund's coupling, irrespectively of the number of orbitals one encounters a zone of negative compressibility\cite{demedici_compress}. This is illustrated in the upper panels of Fig. \ref{fig:compress} where the chemical potential $\mu$ as a function of the total density $n$ is plotted, for the 2-orbital and the 3-orbital Hubbard model on the Bethe lattice with $J/U=0.25$.
Curves for various values of U are plotted. The lowest value illustrates a case in which at half filling a metallic solution is obtained, just below $U_c$ for the Mott transition. All other values of U lead to a Mott insulator at half-filling. A clear change of behavior of the $\mu$ vs n curve appears for $U>U_c$. Indeed upon approaching the Mott insulator the electronic compressibility $\kappa_{el}={dn}/{d\mu}$ diverges (i.e. $\mu(n)$ has a flat slope). At large doping the slope is positive signaling a stable phase, while in a zone near the Mott insulator the negative slope signals negative compressibility and thus an unstable electronic fluid. In the 2-orbital model for all dopings below the one where the divergence happens the system is unstable, giving rise to the purple instability zone in the phase diagram depicted in the lower panel of Fig. \ref{fig:compress}. In the 3-orbital (Fig. \ref{fig:compress}, right panel) and in the 5-orbital model (not shown, see supplementary material in Ref. \cite{demedici_compress}) instead the compressibility becomes positive again before the filling reaches half. This gives rise to a different shape of the instability zone (respectively in green and light blue for the 3- and 5-orbital model in the lower panel of Fig.  \ref{fig:compress}) that appears more like a "moustache".

The striking feature is that in all cases the instability zone departs from the Mott transition point at half-filling (symbols in the lower panel of Fig.  \ref{fig:compress}). 
It is in fact easy to verify that the lower frontier between the stable and unstable metal approximatively coincides with the crossover into the Hund's metal that we have described in this chapter. 
Indeed this is illustrated in the upper panels in Fig. \ref{fig:crossover}, which plot quantities at constant density as a function of U along the scans of the phase diagram signaled in the lower panel of Fig.  \ref{fig:compress} by dashed lines. The (inverse) electronic compressibility $\kappa_{el}^{-1}$ is plotted and the instability is signaled by its vanishing. It is easy to see that for the scans crossing the frontier the compressibility divergence happens in correspondence (or immediately beyond) the crossover into the Hund' metal phase. It is also worth mentioning that in the proximity of the instability zone the compressibility of the stable Hund's metal remains enhanced. In particular this is shown from the scans reported in Fig. \ref{fig:crossover} at larger dopings in both models: indeed even without diverging, the compressibility is strongly enhanced (i.e.  $\kappa_{el}^{-1}$ is very small) in a zone starting with the Hund's crossover.

The interest of this finding lies in the connection between enhanced or diverging compressibility and superconductivity.\index{enhanced compressibility and superconductivity}
Indeed the compressibility being the uniform and static charge-charge response function, its divergence signals an instability of the system towards phase separation (or more physically towards a charge density wave, when taking into account the long range Coulomb interaction neglected in the Hubbard model). A second order transition due to a phase separation instability is a possible cause of Cooper pairing through quantum critical fluctuations in its proximity. Moreover in a Fermi-liquid phase such as the one described in the SSMF the compressibility reads\cite{nozieres}\footnote{The formula reported here (where for instance the renormalized and bare densities of states are simply related by  $N^*(\eps_F)=N(\eps_F)/Z$, - with $\eps_F$ the bare Fermi energy) holds only in simple cases (single band, degenerate identical bands). In the general case, however, proper generalizations can be made, and the considerations made in this section hold valid.}:\index{Fermi liquid}
\be\label{eq:kappa_FL}
\kappa_{el}=\frac{N(\eps_F)/Z}{1+F_0^s}
\ee 
where $N(\eps_F)$, the total bare density of states at the fermi energy, coincides with the non-interacting compressibility. $F_0^s$ is the isotropic, spin-symmetric Landau parameter. It is seen that when Z behaves smoothly as in our case, the divergence (or the strong enhancement) is due to a negative Landau parameter approaching the value $F_0^s=-1$.
The Landau parameter embodies the effect of quasiparticle interactions and a negative $F_0^s$ signals an attraction in the particle-hole channel, which is known to favor superconductivity.

Moreover the enhanced compressibility signals also the enhancement of some quasiparticle interaction vertices with bosonic excitations (like phonons for instance).
Indeed for example for the density-vertex $\Lambda(q,\omega)$, which plays a role in the interaction between electrons and phonons, the following Ward identity holds\cite{Grilli_El-Ph}:
\be
\Lambda(q\rightarrow 0, \omega=0)=\frac{1/Z}{1+F_0^s}.
\ee
One can see that this vertex is enhanced in the same way as the compressibility.
This enhancement of quasiparticle-boson interactions can be a source of instabilities and in particular of an instability in the Cooper channel, i.e. superconductivity.

One word is in order on the possible mechanism causing such a compressibility enhancement in correspondence of the Hund's metal crossover.
We will only give a plausibility argument, that needs to be verified in further studies. 
A doped Mott insulator typically has a shifted spectrum, compared to the insulating case, in which the chemical potential has "jumped into" one of the Hubbard bands. At low electron doping it lies typically at the bottom (and symmetrically, in case of hole-doping) of this band. Now we have mentioned that the width of the Hubbard bands, albeit in a M-orbital system orbital fluctuations would favor a value $\sim\sqrt{M}W$,  is brought back to values of order W by Hund's coupling, near half-filling. However the mechanism behind this shrinking that we have described in the previous section holds only near half-filling, and it is natural to expect that with doping it becomes gradually ineffective, and the width of the Hubbard bands is brought back to $\sim\sqrt{M}W$. Now if the Hubbard band swells with increasing doping, the situation in which the chemical potential at a higher density has a lower value than at a smaller density can naturally happen. This indeed coincides with a negative electronic compressibility.

\begin{figure}[h!]
\begin{center} 
  \includegraphics[width=4.25cm]{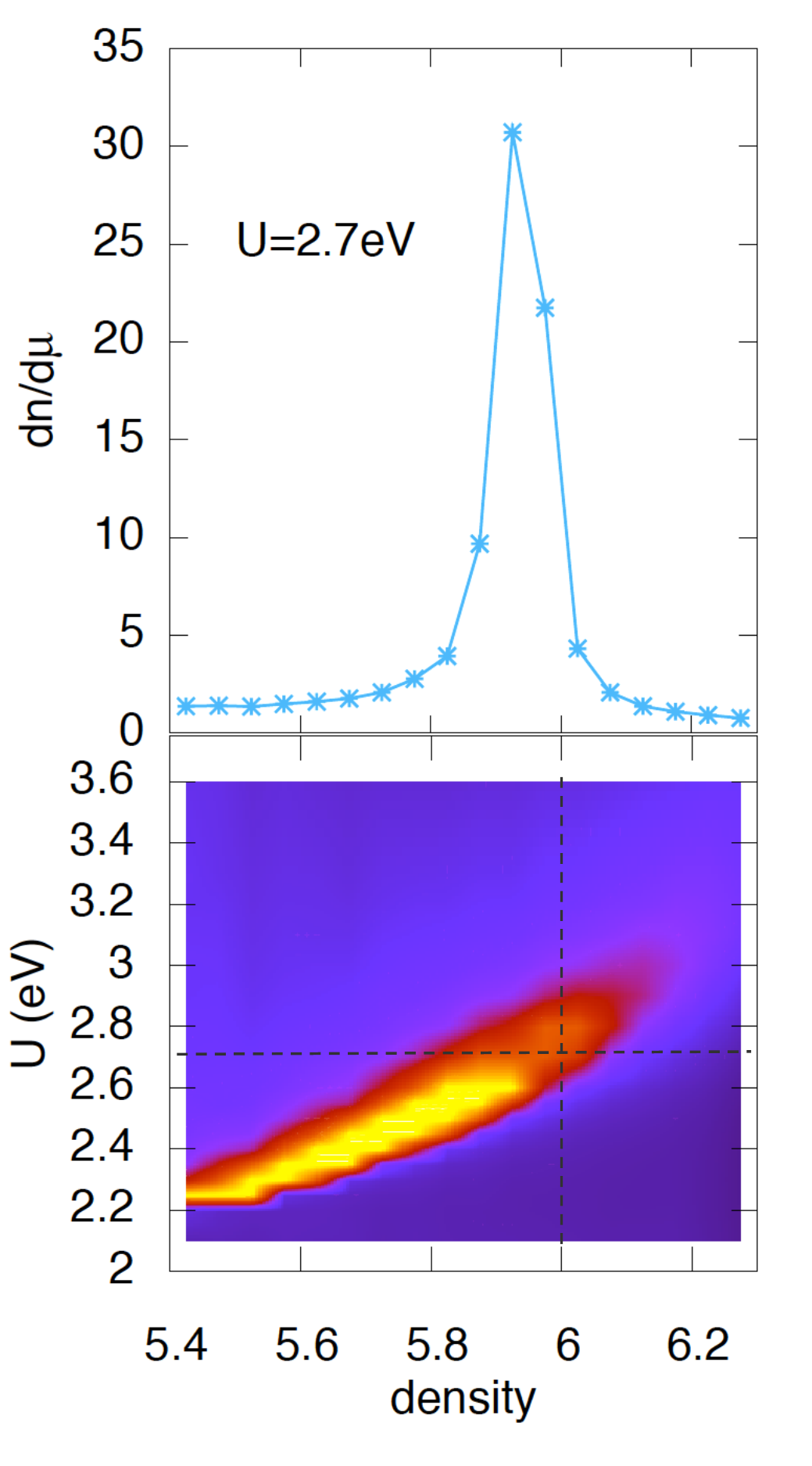}  
   \includegraphics[width=4.25cm]{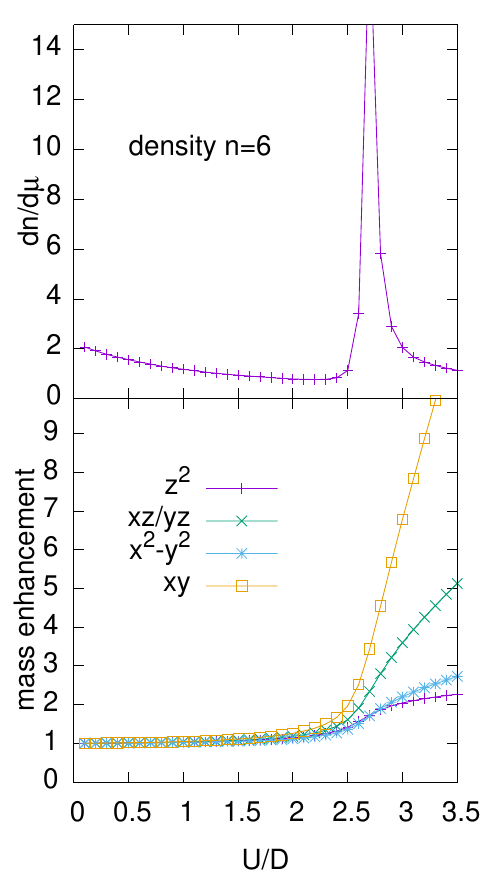}  
  \caption{Lower left panel: compressibility (color scale) in BaFe$_2$As$_2$ calculated within DFT+SSMF (J/U=0.25) in the U-density plane. The saturated yellow color corresponds to the unstable region (the pixelation is due to numerical discretization and is unphysical), surrounded by an area of enhanced compressibility (red). Upper panels: compressibility plotted along the cuts (dashed lines) for constant U=2.7eV and constant density n=6, relevant values for BaFe$_2$As$_2$. Lower right panel: orbitally resolved mass enhancements, showing that the compressibility enhancement happens near the Hund's metal crossover, signaled by the correlations become orbitally selective.}
  \label{fig:BaFe2As2}
\end{center}
\end{figure}

Finally it is worth going back to realistic calculations in FeSC to see if the compressibility enhancement too survives in an ab-initio framework.
In Fig. \ref{fig:BaFe2As2} we report the calculation done within DFT+SSMF for BaFe$_2$As$_2$. The "moustache" expected in the 5-orbital model is well visible and actually goes through the realistic values for the interaction expected in this material (U=2.7eV and J/U=0.25) and used in all the calculations reported in this chapter that successfully reproduce the phenomenology of the 122 family. The moustache (or its prolongation where the divergence becomes a strong enhancement) goes through the right interaction strength exactly for the zone of densities where the material shows both the superconductive and the magnetic instabilities. From the plotted panels it is obvious that the enhancement once again happens concomitantly with the crossover in the Hund's metal regime, signaled by the onset of orbital selectivity of the masses in the figure.

This is a possible confirmation of the role played by the enhancement of the quasiparticle interaction signaled by the compressibility peak in promoting the instabilities in general and maybe high-Tc superconductivity in particular.

\section{Conclusions}

In this chapter we have adopted a clear-cut definition of a Hund's metal, as a metallic phase emerging clearly in the phase diagram of Fe-based superconductors, signaled by three main features: enhanced electron masses, local magnetic moments and orbital-selectivity, and all growing with hole doping (in the case of these materials where the conduction bands are filled more than half). 

We have shown that experiments and theoretical simulations within density functional theory + slave-spin mean field go hand in hand, pointing to the local electronic correlations triggered by Hund's coupling as the origin of this phenomenology.

Then we have shown that all features of the Hund's metal are found also in Hubbard models with featureless band structures, thus proving the generality of this physics and its robustness by respect to details of the materials. The only necessary condition is to be under the influence of a half-filled Mott insulator in presence of sizable Hund's coupling. A crossover line between the normal and the Hund's metal exist, where all the aforementioned features become enhanced, it departs from the Mott transition point at half-filling and extends in the interaction/doping plane.
 
We have then given some analytical arguments in order to gain insight into this phenomenology, showing how Hund's coupling favors the realization of a half-filled Mott insulator at values of the interaction less than the bandwidth, and a mechanism lying at the basis of the "orbital decoupling" causing orbital-selective correlation strength. More on this can be found in Refs. \cite{georges_annrev,demedici_Vietri}.

We have finally pointed out a relatively recent extra feature of the normal to Hund's metal crossover, which is the enhancement of the electronic compressibility. This culminates in a zone of instability towards phase separation, departing from the Mott transition point at half-filling and following the Hund's metal frontier. This phenomenology is common, once again, to models and realistic simulations, and it stems from the outlined Hund's local correlation physics. A Fermi-liquid analysis highlights the connection with an enhancement of quasiparticle interactions, tracing a possible, hitherto unsuspected link between Hund's physics and high-Tc superconductivity. More on this can be found in Ref. \cite{demedici_compress} and in its supplementary material.

\vspace{1cm}
\subsubsection*{Acknowledgements}
The author is supported by the European Commission through the ERC-StG2016, StrongCoPhy4Energy, GA No724177.

%%%%%%%%%%%%%%% References %%%%%%%%%%%%%%%%%
\clearpage
%\bibliographystyle{correl}
%\bibliography{Bib/publdm,Bib/bibldm,Bib/hund,Bib/Janus,Bib/biblio,Bib/FeSC}

\clearchapter

\end{document}